\documentclass[dvips]{article}

\usepackage{icrc2011}

\title{The Potential of Spaced-based High-Energy Neutrino Measurements via the Airshower Cherenkov Signal}

\shorttitle{J.F. Krizmanic \& J.W. Mitchell: Space-based, neutrino-induced Airshower Cherenkov Detection}

\authors{John F. Krizmanic$^{1,2}$, John W. Mitchell$^{2}$ }
\afiliations{$^1$CRESST/USRA\\ $^2$NASA/Goddard Space Flight Center, Greenbelt, Maryland, 20771  USA }
\email{John.F.Krizmanic@nasa.gov}

\abstract{Future space-based experiments, such as OWL \cite{OWL} and JEM-EUSO \cite{EUSO}, view large atmospheric and terrestrial neutrino targets.  With energy thresholds slightly above $10^{19}$ eV for observing airshowers via air fluorescence, the potential for observing the cosmogenic neutrino flux associated with the GZK effect is limited.  However, the forward Cherenkov signal associated with the airshower can be observed at much lower energies.  A simulation was developed to determine the Cherenkov signal strength and spatial extent at low-Earth orbit for upward-moving airshowers. A model of tau neutrino interactions in the Earth was employed to determine the event rate of interactions that yielded a tau lepton which would induce an upward-moving airshower observable by a space-based instrument. The effect of neutrino attenuation by the Earth forces the viewing of the Earth's limb to observe the $\nu_\tau$-induced Cherenkov airshower signal at above the OWL Cherenkov energy threshold of $\sim$$10^{16.5}$ eV for limb-viewed events.  Furthermore, the neutrino attenuation limits the effective terrestrial neutrino target area to $\sim$$3 \times 10^{5} km^2$ at $10^{17}$ eV, for an orbit of 1000 km and an instrumental full Field-of-View of 45$^\circ$.   This translates into an observable cosmogenic neutrino event rate of $\sim$1/year based upon two different models of the cosmogenic neutrino flux, assuming neutrino oscillations and a 10\% duty cycle for observation.}
\keywords{ Neutrino, Space-based Measurements, UHECRs, Cherenkov, Simulations}

\begin{document}
\maketitle


\section{Introduction}
\vspace{-1mm} 
Future space-based air fluorescence experiments employ wide field-of-view optics from a orbiting platform(s) to monitor a vast amount of the atmosphere.  For the OWL (Orbiting Wide-angle Light Collectors) mission, the mass of the viewed atmosphere corresponds to more than $10^{13}$ metric tons (mtons). The design choices for OWL were driven by the goal to measure the UHECR spectrum, via the air fluorescence technique, with high statistics above $\sim$$10^{19}$ eV. Studies indicated that the ability to measure neutrino interactions in the atmosphere via the air fluorescence signature exists, but the predicted event rate based upon cosmogenic neutrino flux models \cite{Bartol} is $< 1$/year (assuming a duty cycle of $10$\%) due to the paucity of neutrino flux above $10^{19}$ eV.  Furthermore, the neutrino event rate quickly diminishes as the energy threshold becomes further away from the $\sim$$3 \times 10^{19}$ eV threshold for full neutrino aperture, which assumes both OWL satellites stereoscopically view each event.

Airshowers also produce an intense, beamed Cherenkov signal. OWL simulation studies indicated that the energy threshold for observing the optical Cherenkov signal form a nadir-viewed, upward-moving vertical airshower initiated by a particle at sea level would be $\sim$$10^{15.5} eV$. OWL also views a large terrestrial area: assuming 1000 km orbits and the two OWL satellites are tilted to view a common area, the terrestrial area monitored ranges from $6 \times 10^{5}$ km$^2$ for a 500 km satellite separation to nearly $2 \times 10^{6}$ km$^2$ for a 2000 km satellite separation.  These vast areas offer a large neutrino target, depending upon the depth that provides a measurable signal.  Tau neutrino charged-current (CC) interactions offer a mechanism to exploit this large, terrestrial neutrino target: at high energies, the produced tau lepton has a sufficient Lorentz-boosted length to escape the Earth, decay in the atmosphere, and create an upward moving airshower that could be observed via the Cherenkov signal.

This paper details the calculations used to quantify the sensitivity of measuring the cosmogenic neutrino flux using space-based measurements, assuming the performance defined by the OWL experiment, of the Cherenkov signal created from upward-moving tau neutrino induced airshowers originating in a terrestrial neutrino target.

\vspace{-2mm} 
\section{Optical Cherenkov Signal Simulation}
\vspace{-1mm} 
In order to determine the optical Cherenkov signal strength and profile at a orbiting instrument, a computer-based simulation was constructed based upon parameterizations described by Hillas \cite{Hillas1, Hillas2}. The charged-particle density for an airshower is given by the Greisen parameterization as a function of shower age, with the airshower electron angular distribution and the energy-dependent, charged-particle track length fraction defined by well-behaved mathematical functions dependent upon shower age.  The index of refraction of air is given as a function of atmospheric grammage, or altitude, which defines the Cherenkov angle and Cherenkov energy threshold. The latter, combined with the charged-particle track length fraction, determines the fraction of the particles in the airshower that contribute to the Cherenkov signal for a given altitude. The atmosphere is described by the Shibata parameterization \cite{Gaisser}, and the wavelength-dependent attenuation of UV light in the atmosphere is described by a parameterization \cite{Krizmanic} based upon more detailed models.

\begin{figure}[!t]
  \centering
  \includegraphics[width=2.75in]{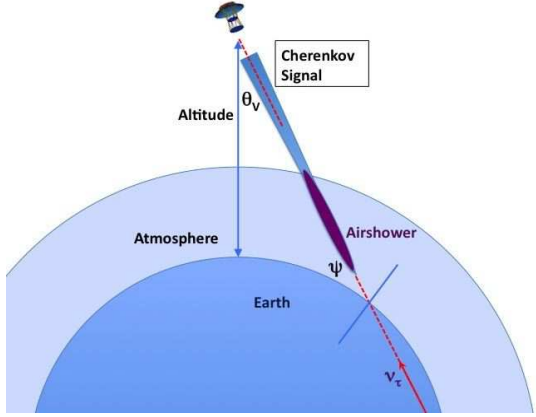}
  \caption{Schematic of the Cherenkov signal generated by an upward-moving airshower induced by a tau neutrino interacting in the Earth.}
  \label{figA}
\vspace{-2mm}
 \end{figure}

The parameterizations are then used to generate the Cherenkov signal at an arbitrary altitude for an upward-moving airshower. The lateral spread of the charged particles in the shower was not considered since this is a small effect for large viewing distances.  A 100 m airshower step size was used starting at sea level, and the sampling of the electron angular distribution at each step was $\le 10^{-3}$ radians. The charged particle track length fractions were sampled in decades of energy for each step, from the Cherenkov energy threshold  to a factor of 10 below the total airshower energy.  The Cherenkov light was generated at each step from 200 nm to 600 nm in increments of 25 nm, but the effects of ozone absorption, in the atmospheric attenuation model, significantly reduces the signal below $\sim$300 nm.  Two numerical azimuthal integrations were performed: one defined by the Cherenkov angle about the electron angular sampling vector, and the other by angular sampling vector about the shower direction.

The effects of the Earth's curvature were modeled as these become important for viewing angles away from the nadir direction.  Figure 1 illustrates the geometry of this effect in relation to an upward moving airshower induced by a tau neutrino interaction in the Earth near the Earth's limb. The Cherenkov airshower simulation accounts for this effect by determined the proper angle, shown as $\psi$ in the figure, at each shower propagation distance step, relative to the viewing angle $\theta_V$.  Thus the atmospheric attenuation depth and the distance from the shower and the UV detector in orbit can be accurately determined.

 \begin{figure}[!t]
 \vspace{-9mm}
  \centering
  \includegraphics[width=2.9in]{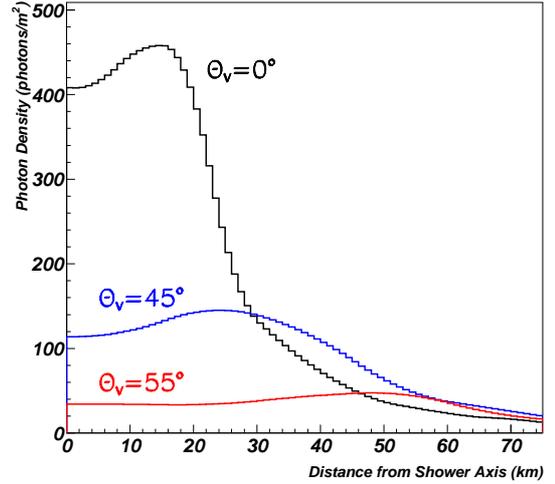}
  \caption{The radial profile of the simulated Cherenkov light density at an altitude of 1000 km for upward-moving $10^{17}$ eV airshowers viewed at angles of 0$^\circ$, 45$^\circ$, and 55$^\circ$.}
  \label{figB}
\vspace{-2mm}
 \end{figure}

The results of the Cherenkov airshower simulation are shown in Figure 2 for $\theta_V=0^\circ, 45^\circ,$ and $55^\circ$ at an altitude of 1000 km. The curves show the Cherenkov light density (photons/m$^2$) as a function of radial distance from the projected location of the airshower core in the plane perpendicular to the airshower direction at 1000 km altitude. The form of the curves show that the Cherenkov light cone (illustrated in Figure 1), is approximately uniform in photon density up to a radius which closely corresponds to that defined by the Cherenkov angle at shower maximum and the distance from shower maximum to the measurement.  This implies that a good energy resolution can be achieved by sampling the uniform part of the distribution, e.g. by one of the two OWL satellites. The lateral size of the Cherenkov light cone also defines the solid angle: $1.3 \times 10^{-3}$ steradians for the nadir ($\theta_V = 0^\circ$) case and $1.9 \times 10^{-3}$ sr for $\theta_V=55^\circ$, which corresponds to viewing close to the Earth's limb (at an altitude of 1000 km, the Earth's limb is given by $\theta_V \approx 60^\circ$). A comparison of the results presented in Figure 2, after scaling the energy and altitude, are in good agreement (30\% difference) with results which employ a full airshower Monte Carlo and more detailed atmospheric attenuation modeling \cite{Kieda}. 

The factor of $\sim$10 decrease in the photon density for $\theta_V=55^\circ$, as compared to the nadir case, is mainly due to the effect of the Earth's curvature.  A factor of $\sim$5 reduction is due to a $1/r^2$ effect (for $0^\circ$ the distance from 1000 km altitude to shower max is 990 km while it is 2190 km for the $55^\circ$ case) and a factor of $\sim$2 reduction is due to atmospheric attenuation of the Cherenkov signal.

The signal strength in an OWL 'eye' is defined by the 3 m optical aperture, the optical transmission, and the quantum efficiency of the focal plane detector.  When these are combined, the photo-electron signal strength in an OWL pixel is approximately the photon density in units of photons/m$^2$, which unit of the ordinate axis in Figure 2.  Thus a Cherenkov signal of 400 photons/m$^2$ corresponds to a signal in a OWL instrument of $\sim$400 photo-electrons.

\vspace{-2mm} 
\section{Cosmogenic Neutrino Rate Determination}
\vspace{-1mm} 
For charged-current tau neutrino interactions in the Earth, the target depth is estimated by the propagation distance of the tau lepton: $\gamma c \tau$.  At energies above $10^{16}$ eV, neutrinos and antineutrinos have similar cross-sections for neutrino-quark scattering, and the produced lepton carries $\sim$75\% of the incident neutrino energy at $10^{16}$ eV with the energy fraction increasing to $\sim$80\% at $10^{20}$ eV \cite{NuCross}.  $E_\tau = E_\nu$ is assumed for this ICRC paper.  At higher energies, catastrophic energy losses (a convenient parameterization is used \cite{Weiler}) limit the effective depth of the neutrino target above $\sim$$10^{18}$ eV.  Table 1 details the terrestrial neutrino target depth for $\nu_\tau$ interactions, as a function of $E_\nu$, for both water and rock and calculates the mass for a target area of $10^6$ km$^2$ assuming rock ($\rho = 2.3$ g/cm$^3$). Note that tau energy losses limit the target depth above $\sim$$10^{18}$ eV while the maximum neutrino target mass also occurs at $\sim$$10^{18}$ eV.

\begin{table}[t]
\begin{center}
\begin{tabular}{l|ccc}
\hline
$E_\tau$  &  Depth (km) &   Depth (km) &  Mass (mtons) \\
 (eV) &  ($\rho =1 \frac{g}{cm^3}$) &  ($\rho =2.3\frac{g}{cm^3}$)       &  for 10$^6$ km$^2$ \\
\hline
$10^{15}$ & 0.05  & 0.05  & $1.3 \times 10^{14}$   \\
$10^{16}$  & 0.5  & 0.5 & $1.3 \times 10^{15}$   \\
$10^{17}$    & 5 & 5  & $1.3 \times 10^{16}$   \\
$10^{18}$   & 29  & 16   & $7.6 \times 10^{16}$   \\
$10^{19}$   & 18  & 10   & $4.8 \times 10^{16}$   \\
\hline
\end{tabular}
\caption{The terrestrial neutrino target depth for $\nu_\tau$ CC interactions, for water and rock targets, and the target mass for $10^6$ km$^2$ for rock as a function of energy.}
\label{tableA}
\end{center}
\vspace{-5mm}
\end{table}

Assuming a terrestrial neutrino target area of $10^6$ km$^2$, the rate of cosmogenic neutrino interactions that lead to an observable tau-induced, upward-moving airshower via the Cherenkov signal can be calculated by numerically integrating the neutrino flux, cross-section,  target mass, $\ldots$, with respect to the neutrino energy, and accounting for the solid angle of the Cherenkov signal.  Two cosmogenic flux models were used, the Bartol ($\Omega_\Lambda=0.7$) model \cite{Bartol} and a more recent calculation by Scully\&Stecker \cite{SS}. The models, both which assume proton dominated UHECRs, provide the electron and muon neutrino and antineutrino fluxes. These were summed for each model and divided by 3 to predict the $\nu_\tau$ flux, eg assuming neutrino oscillation yields equal flavor fractions at Earth.

 \begin{figure}[!t]
 \vspace{-9mm}
  \centering
  \includegraphics[width=3.0in]{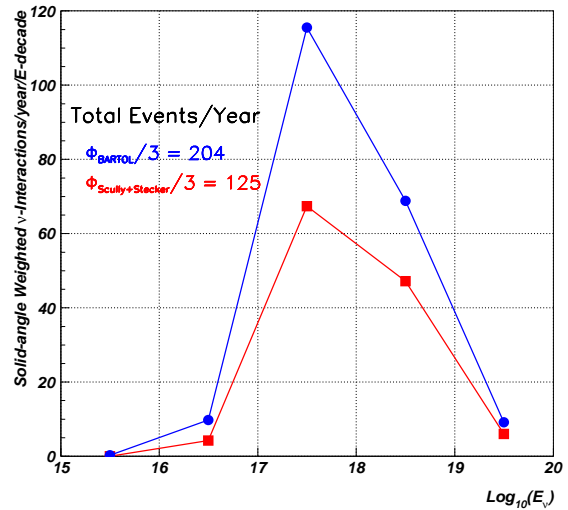}
  \caption{The rate of observable tau neutrino interactions via the airshower Cherenkov signal for two cosmogenic neutrino flux models. The rate is presented in number of events/year per energy decade.}
  \label{figB}
\vspace{-2mm}
 \end{figure}

Figure 3 shows the results presented as observable events/year per energy decade for the two cosmogenic models, assuming a solid angle of $1.5 \times 10^{-3}$ sr, a terrestrial neutrino target area of $10^6$ km$^2$ and a 100\% duty cycle for the experimental measurement of the Cherenkov signal.  The rate prediction based upon the Bartol flux is given by the upper curve while that based upon the Scully\&Stecker model is given by the lower curve. An interesting feature is that the event rate does not become appreciable until above  $10^{16}$ eV, indicating that the neutrino target mass, and, to a lesser extent, the neutrino cross-sections, becomes sufficiently large at this energy to yield a significant rate.

While the neutrino rate prediction of $> 100$ events/year is encouraging, the calculation does not take into account the fact that the duty cycle for observation of the Cherenkov signal is closer to $10\%$ and the Earth is relatively opaque to these energetic neutrinos.  This latter point forces the orbiting instrument to view the limb of the Earth to observe airshowers induced by neutrinos with sufficiently short path lengths in the terrestrial target.  Figure 4 details the geometry: the orbiting detector must be tilted to view the limb of the Earth, and the effective terrestrial neutrino target is constrained to be no larger than an appropriately short neutrino path length in the Earth.

\begin{table}[b]
\begin{center}
\begin{tabular}{l|cc}
\hline
Altitude (km) &  Full Area (km$^2$) &   Effective Area (km$^2$)  \\
\hline
500  &  $1.7 \times 10^6$ & $1.6 \times 10^5$ \\
1000  &  $3.8 \times 10^6$ & $2.6 \times 10^5$ \\
2000  &  $8.9 \times 10^6$ & $3.9 \times 10^5$ \\
\hline
\end{tabular}
\caption{The full and effective viewed surface areas for an $45^\circ$ FOV instrument, tilted to view the Earth's limb, as a function of altitude. The effective area is constrained by chord equal to the $\nu$ interaction length in Earth at $10^{17}$ eV.}
\label{tableB}
\end{center}
\end{table}

The problem now effectively becomes one of analytic geometry. The tilt angle of the instrument is defined by its FOV and the angle to the Earth's limb, the calculation of the viewed elliptical area on the Earth's surface is determined by the intersection of the tilted cone defined by the FOV and a sphere of radius 6378 km.  The 2-dimensional (eg flat) area of the ellipse can be determined using the conic section relationship between the angle of the plane, which defines the ellipse in the cone, and the eccentricity of the ellipse \cite{cone}.  The 2-dimensional area of the viewed ellipse was then increased by 5\% to account for the effects of the sphere's (Earth's) curvature. The effective area (darkest area in Figure 4) is defined by the chord whose length is the neutrino attenuation length in Earth, for a particular energy.  This describes a truncated ellipse which is inscribed in a rectangle of calculable dimensions. Assuming the truncated ellipse can be approximated by a parabola of similar width, one can use the fact that the area of the parabola is 2/3 that of the bounding rectangle \cite{Arch} to approximate the area of the truncated ellipse, which is the effective area of the terrestrial neutrino target.

Table 2 presents the total, tilted viewed terrestrial area for an instrument with $45^\circ$ FOV and the effective neutrino target area at $10^{17}$ eV, as a function of altitude, determined from the analytic geometrical calculation.  While the full area monitored by a tilted instrument is substantial, the effective neutrino target area available near the Earth's limb is reduced by more than a factor of 10, to $2.6 \times 10^5$ km$^2$, assuming a 1000 km altitude.

Combining the energy-dependent effective area results within the numerical integration using the two cosmogenic flux models and assuming 10\% duty cycle, the predicted observable tau neutrino event rate is reduced to $\sim$2 event/year using the Bartol model and $\sim$1 event/year using that of Scully\&Stecker.

\begin{figure}[!t]
\vspace{-2mm}
  \centering
 \includegraphics[width=2.3in]{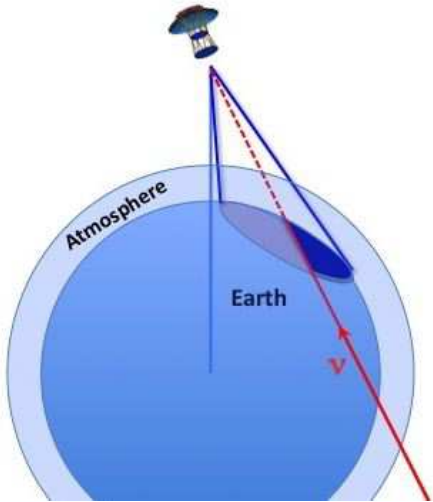}
  \caption{A 2-dimensional schematic of the effective neutrino target area constrained by the $\nu$ interaction length. Only those neutrinos with a chord length sufficiently small will be relatively unattenuated by the Earth.  While a tilted wide FOV UV imager monitors a large, elliptical area of the Earth, only a small (darkest region) portion samples a significant neutrino flux.}
  \label{figD}
\vspace{-3mm} 
\end{figure}

\vspace{-3mm} 
\section{Conclusions}
\vspace{-1mm} 
Simulation studies of the Cherenkov signal from upward-moving airshowers indicate that an orbiting experiment with a 3 m optical aperture and UV sensitivity of an OWL instrument would have an energy threshold of slightly higher than $10^{16}$ eV for airshowers generated near the Earth's limb. This is well-matched for tau-induced airshowers generated by cosmogenic neutrino interactions in the Earth. However, the 10\% duty cycle inherent for the Cherenkov observation and the Earth's attenuation of the neutrino flux limit the effective terrestrial area, which is estimated to be $~\sim$$3 \times 10^{5}$ km$^2$ for $E_\nu = 10^{17}$ eV, assuming an instrument with a $45^\circ$ full FOV tilted to observe the Earth's limb.  Using two different cosmogenic neutrino flux models, the predicted observable event rate is $\sim$1 event/year.  While factor of 2 improvements may be available using different orientations of the two OWL satellites or realizing a gain in the duty cycle, the net effect of these improvements may be balanced by potential decreases caused by more realistic modeling of the energy distribution of the create tau lepton, the airshower generated by the tau decay, and the inherent shower fluctuations. 

While the tau-induced airshowers could be observed via the air fluorescence technique, which has a much larger observational solid angle than that inherent to the Cherenkov signal, the higher energy threshold of $\sim$$10^{19}$ eV (and a factor of $\sim$10 higher for viewing near the Earth's limb), severely limit the sensitivity to the cosmogenic neutrino flux.  This reinforces a similar result from more detailed Monte Carlo studies of the ability of OWL to measure airshowers induced by cosmogenic electron neutrinos in the atmosphere, which predict $<1$ event/year. However, if the energy threshold for air fluorescence could be reduced to $10^{18}$ eV for an OWL-type mission, studies have indicated that the measurable cosmogenic neutrino event rate would be $\sim$50 events/year for the $\nu_e$ atmospheric channel, assuming a 10\% duty cycle.  There could also be a significant observable rate from $\nu_\tau$ interacting in the Earth observed via air fluorescence, if the energy threshold could be reduced.


\clearpage


\vspace{-3mm} 
\begin{thebibliography}{}
\vspace{-1mm} 
\bibitem{OWL} F.W. Stecker et al., NucPhysB, 2004, {\bf 136}: 433-438.

\bibitem{EUSO} JEM-EUSO Collaboration, M. Bertaina, NucPhysB Proc Sup, {\bf 190}: 300-307.

\bibitem{Bartol} R. Engel, D. Seckel, T. Stanev, PhysRevD, 2001, {\bf 64}(9): 093010.

\bibitem{Hillas1} A.M. Hillas, JourPhysG, 1982, {\bf 8}(10): 1461-1473.

\bibitem{Hillas2} A.M. Hillas, JourPhysG, 1982, {\bf 8}(10): 1475-1492.

\bibitem{Gaisser} T. Gaisser:1990, {\it Cosmic Rays and Particle Physics}, Cambridge University Press.

\bibitem{Krizmanic} J.F. Krizmanic, 26th ICRC (Salt Lake): 1999, 2-388.

\bibitem{Kieda} D. Kieda, 27th ICRC (Hamburg): 2001, Paper 6407.

\bibitem{NuCross} Gandhi et al., PhysRevD, 1998, {\bf 58}(9): 093009.

\bibitem{Weiler} S. Palomares-Ruiz, A. Irimia, T. Weiler, PhysRevD, 2006, {\bf 73}(8): id 083003.

\bibitem{SS} S.T. Scully, F.W. Stecker, AstropartPhys, 2011, {\bf 34}(7): 575-580.

\bibitem{cone} www.mamikon.com/USArticles/CircumSolids.pdf.

\bibitem{Arch} Archimedes of Syracuse: 3rd century BC, {\it Quadrature of the Parabola} ; translation via  {\it The Works of Archimedes}: 1953, ed. Sir Thomas L. Heath, Dover.

\end{thebibliography}
\end{document}